\documentclass[aps,prl,reprint,onecolumn,superscriptaddress,nofootinbib]{revtex4-2}

\usepackage[left=1.2in,right=1.2in,top=1in,bottom=1in]{geometry}
\usepackage{amsmath,amssymb,amsfonts}
\usepackage{graphicx}
\usepackage{xcolor}
\usepackage[colorlinks=true,allcolors=green!35!black]{hyperref}
\usepackage{braket}

\begin{document}

\title{Criticality and Quench Dynamics at the Anderson Transition of a Chern Insulator}

\author{L. Ul\v{c}akar}
\email{lara.ulcakar@fmf.uni-lj.si}
\affiliation{ University of Ljubljana, Faculty of Mathematics and Physics, Jadranska 19, Ljubljana, Slovenia}
\affiliation{Jo\v{z}ef Stefan Institute, Jamova 39, Ljubljana, Slovenia}

\author{J. Mravlje}
\affiliation{Jo\v{z}ef Stefan Institute, Jamova 39, Ljubljana, Slovenia}
\affiliation{ University of Ljubljana, Faculty of Mathematics and Physics, Jadranska 19, Ljubljana, Slovenia}

\author{G. Lemut}
\affiliation{Dahlem Center for Complex Quantum Systems, Halle-Berlin-Regensburg Cluster of Excellence CCE, and Fachbereich Physik, Freie Universit\"at Berlin, Arnimallee 14, 14195 Berlin}

\author{T. Rejec}
\affiliation{ University of Ljubljana, Faculty of Mathematics and Physics,  Jadranska 19, Ljubljana, Slovenia}
\affiliation{Jo\v{z}ef Stefan Institute, Jamova 39, Ljubljana, Slovenia}

\date{\today}

\begin{abstract}
We study the critical properties and non-equilibrium dynamics of the topological Anderson phase transition in a strongly disordered Chern insulator. We show that a critical length scale emerges in the real-space profile of the local Chern marker and matches the corresponding correlation-length critical exponent found in the literature. We further determine the dynamical critical exponent from the frequency-dependent conductivity, finding values consistent with those of non-interacting models in the integer quantum Hall universality class.
We then ramp the disorder strength across the phase transition and investigate the ensuing non-equilibrium dynamics. In contrast to clean topological systems, the spectral gap is closed due to localized states and consequently the generation of excitations does not exhibit Kibble--Zurek scaling. Despite this, we find that the non-equilibrium length scale extracted from the local Chern marker shows signatures of the Kibble--Zurek scaling.
Our results demonstrate that the non-equilibrium dynamics of the topological Anderson transition features decoupling between the excitations and the emergent length scale as known from clean systems, and establish the local Chern marker as a probe of the emergent length scale during the quench.
\end{abstract}

\maketitle

\section{Introduction}
\label{sec:intro}
Chern insulators are renowned for the presence of robust electronic edge states and the quantum anomalous Hall (QAH) effect. These properties are a consequence of a non-trivial band topology, which protects them against dissipation \cite{Hasan10,Chang13,Otrokov19,Serlin20}. Numerical studies have shown however that strong enough disorder in the bulk can destroy the topological phase due to Anderson localization~\cite{Prodan10,Medvedyeva10}.
The phase transition into a trivial Anderson insulator is of second order and is signaled by the vanishing of the Chern number~\cite{Mildner23,Lewenkopf24,Bera24,Salib25}. Previous numerical studies of disordered Chern models have shown that the transition is consistent with the universality class of the integer quantum Hall effect (IQHE)~\cite{Chalker88,Huckestein90,Ludwig94}, with the localization length diverging with a critical exponent $\nu\approx2.58-2.62$ ~\cite{Slevin09,Obuse12,Puschmann19,Sbierski21,MorenoGonzalez23}. These findings were based on scaling analyses of the Chern number, the direct-current conductivity and transfer-matrix localization lengths. Here, we show that this critical scaling also appears as the size of inhomogeneous regions in the real space profile of the local Chern marker (LCM), which was observed to fluctuate in Refs.~\cite{Ulcakar20,Favata25}. Such formulation of the length scale can be beneficial, as it remains meaningful also out of equilibrium, as opposed to finite-size scaling analysis performed in the ground state. Through the scaling of the frequency-dependent conductivity we also determine the dynamical critical exponent $z=2$, which was so far unexplored in disordered Chern insulators.

Similarly, non-equilibrium dynamics following quenches across transitions between distinct Anderson-localized phases has not been studied yet.
Quench dynamics has been extensively studied in topological band insulators~\cite{Caio2015,Mitra16,Caio2016,Unal16,zoller,Privitera16,Wilson16,Wang16,Duta17,Schuler17,Ulcakar2018,Ulcakar19,McGinley18,McGinleyPRL18,Liou18,Sim22}. These works showed that even slow quenches generally drive the system out of equilibrium due to the closing of the band gap at the critical point. In translation-invariant systems, the dynamics near the gap closing can be described by Landau–Zener band crossing, which is directly connected to the Kibble–Zurek (KZ) mechanism~\cite{Kibble,Zurek}. 
This non-equilibrium paradigm describes the dynamics as adiabatic except close to the critical point, where it is effectively frozen. Consequently, the non-equilibrium length scale and its associated topological defects inherit the properties of the critical point reflected in the scaling determined by the critical exponents. In translation-invariant topological systems, the excitation density and characteristic length scale obey universal KZ scaling relations~\cite{Damski05,Dutta10,Ulcakar2018,Ulcakar19}. 
Recent studies further demonstrated that weak disorder induces spatial inhomogeneities in real-space topological markers that are characterized by a length scale consistent with KZ predictions~\cite{Ulcakar20,Sun22,Yuan24}. 
In all these cases, it is the  single gap closing that ties both the excitation density and the non-equilibrium length scale to the equilibrium critical exponents.

At strong disorder this is no longer the case. In this work, we show that the spectral gap of the disordered Chern insulator is filled with Anderson-localized states already at a disorder strength well before the topological Anderson phase transition.  When the disorder strength is varied in time across this transition, the localized in-gap states undergo a large number of avoided crossings at the Fermi energy and are therefore excited by any finite-rate ramp \cite{Altshuler10}. Therefore, the Landau-Zener description of collective band crossings, on which the KZ mechanism rests, does not apply. Our results show that the resulting excitation density exhibits only a weak power-law dependence on the quench time and deviates from the KZ prediction. Despite the absence of adiabaticity, we find that the non-equilibrium LCM length scale exhibits signatures of the freeze-out behaviour and the KZ scaling. This occurs because the LCM is sensitive to the topological transition rather than to the closing of the spectral gap.

The paper is structured as follows. In Sec.~\ref{sec:model} we introduce the model and methods. In Sec.~\ref{sec:PH} we inspect the topological properties through the LCM, the electrical conductance and the spectral statistics. In Sec.~\ref{sec:crtitical}, we extract the dynamical and the correlation length critical exponents from the frequency-dependent conductivity and the LCM, respectively. In Sec.~\ref{sec:quenches}, we present our main results. We study finite-time quenches of the disorder strength across the transition, starting from the ground state in either the topological phase or the trivial Anderson localized phase. The non-equilibrium profiles of the LCM and the density of excitations are analysed and discussed. In Sec.~\ref{sec:conclusion}, we conclude the paper.

\section{The model and methods}
\label{sec:model}
We study an Anderson localization phase transition in a Chern insulator described by the  half-filled disordered Qi-Wu-Zhang (QWZ) model \cite{QWZmodel} 
\begin{equation}
\hat H=\sum_{\mathbf r}|\mathbf r\rangle\langle \mathbf r|\otimes  (u\,\hat\sigma_z+W(\mathbf{r})\,\hat\sigma_0)
+\sum_{\mathbf r,j\in\{x,y\}}\left(|\mathbf r + \boldsymbol{e_j}\rangle\langle \mathbf r|\otimes\tfrac{\hat\sigma_z+i\hat\sigma_j}{2} +\textrm{h.c.}\right)
\label{eq:QWZreal}
\end{equation}
where $\mathbf{r}=(x,y)$ are the Bravais lattice vectors of a square lattice, measured in units of the lattice constant, and $\boldsymbol{e_j}$ are Cartesian unit vectors. The system size is $N\times N$ unit cells and periodic boundary conditions are assumed. Each unit cell hosts two orbitals $|\mathbf{r}\alpha\rangle$, $\alpha\in\left\{A,B\right\}$, and the Pauli matrices $\hat\sigma_i$, $i\in\{x,y,z\}$, act on these orbital degrees of freedom. $W(\mathbf{r})$ is the electrostatic disorder that is uncorrelated and uniformly distributed on the interval $[-W/2,W/2]$. In the following, the system size is set to $N=100$ and results are averaged over $100$ disorder realizations, unless explicitly specified otherwise.  The QWZ model has been realized with ultracold atoms~\cite{Liang23}. The following results obtained for the QWZ model can be generalized to the whole class of models for two-dimensional Chern insulators, as they belong to the same universality class.

We characterize the topological phases of inhomogeneous states by calculating the LCM. The LCM is a real-space analogue of the Berry curvature and was shown to reflect critical behaviour of the topological phase transitions~\cite{Caio19,Molignini23} in  disordered systems~\cite{Ulcakar20,Favata25}. It is defined as~\cite{Cmarker}
\begin{equation}
c(\mathbf{r})=2\pi i \sum_{\alpha}\langle  \mathbf{r}\alpha|\hat P[-i[\hat x,\hat P],-i[\hat y,\hat P]]|\mathbf{r}\alpha\rangle,
\label{LCM}
\end{equation}                
where $\hat P=\sum_{n}|\Psi_n\rangle\langle\Psi_n|$ is the projector onto the subspace spanned by occupied states $|\Psi_n\rangle$. For a system with periodic boundary conditions, $-i[\hat x,\hat P]$ is expressed with finite differences as proposed in Ref.~\cite{Prodan10}. The Chern number is then $C=\lim_{N\to\infty}\frac{1}{N^2}\sum_\mathbf{r}c(\mathbf{r})$ \cite{Cmarker}. 

We additionally characterize the state by examining the two-terminal conductance $G$ of the system using the KWANT package~\cite{Kwant}. Periodic boundary conditions are assumed in the other direction. The electrical conductance is calculated from the transmission matrix $\hat{t}(E)$ for scattering states at energy $E$ as $G(E)=G_0\,\mathrm{tr}(\hat{t^\dagger}(E)\hat{t}(E))$, where $G_0=e^2/h$ is conductance quantum, $h$ is Planck constant and $e$ is the electron charge.

In order to inspect the dynamical critical properties, we also calculate the linear-response alternating-current (AC) conductivity in the limit of linear response. It characterizes the response of the system to an oscillating electrical field with frequency $\omega$. Its real part is expressed as:
\begin{equation}
\sigma_{xx}(\omega)=-\lim_{\eta\to 0}\frac{e^2\hbar}{N^2}\sum_{m\neq n}\frac{f(E_m)-f(E_n)}{E_m-E_n}|\langle \Psi_m|\hat{v}_x|\Psi_n\rangle|^2\frac{\eta}{(E_m-E_n-\hbar\omega)^2+\eta^2}.
\end{equation}
Here, $\hat{v}_x=i/\hbar[\hat H, \hat x]$ is the velocity operator in $x$-direction, $f(\varepsilon)$ is the Fermi-Dirac occupation function, and $E_n$ and $|\Psi_n\rangle$ are the eigenenergies and eigenstates, respectively. 
For numerical simulations of finite systems, $\eta$ should be of the order of the energy level spacing. We set it to $\eta=15/N^2$.

\section{Phase Diagram}
\label{sec:PH}
\begin{figure}[h]
\includegraphics[width=0.325\textwidth]{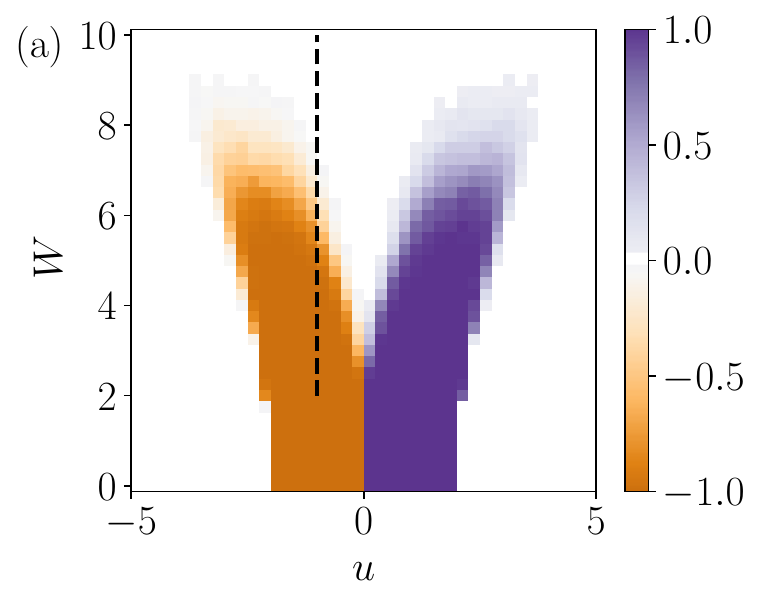}
\includegraphics[width=0.31\textwidth]{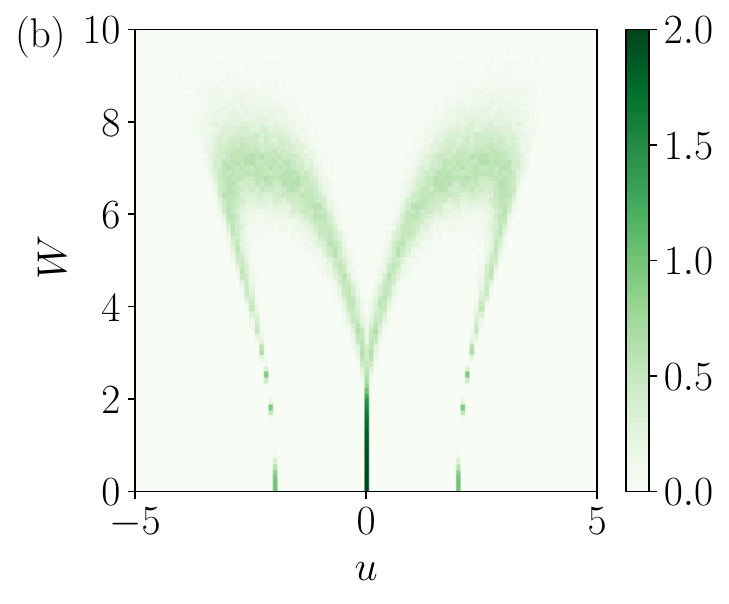}
\includegraphics[width=0.325\textwidth]{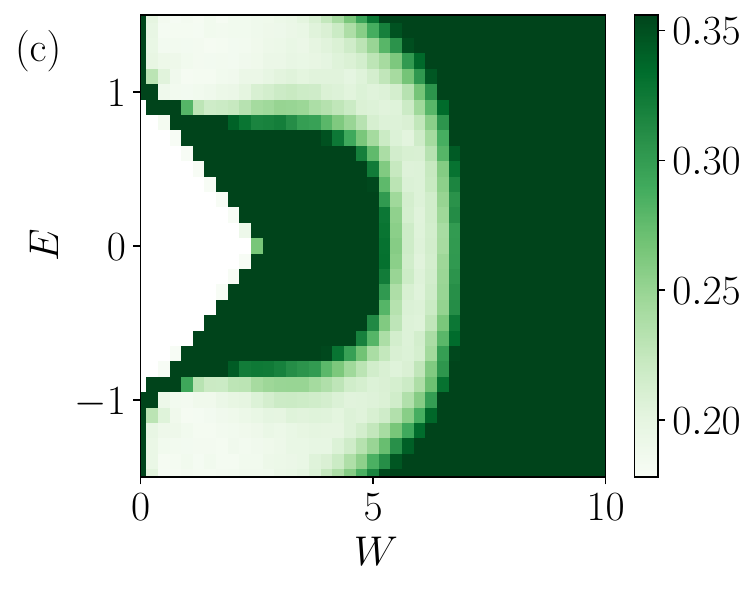}
	\caption{Phase diagram in the $(u, W)$ parameter space of the (a) Chern number and (b) the electrical conductance. (c) The energy level-spacing variance of the states per energy window of width $0.1$,  at $u=-1$. Data is averaged over (a) $10$, (b) $50$ and (c) $100$ disorder realizations. Dashed line indicates the values of parameters during the quench: $u$ is fixed to $-1$ and $W$ is varied between $2$ and $10$.}
	\label{fig:PD}
\end{figure}

In the absence of disorder, the system is in a topologically non-trivial insulating phase with $C=\mathrm{sgn}(u)$ for $|u|<2$ and a trivial insulating phase for $|u|>2$. As shown in Fig.~\ref{fig:PD}(a) and demonstrated in previous works~\cite{Prodan10,Mildner23,MorenoGonzalez23,Bera24,Salib25}, the topologically non-trivial phase persists until large values of disorder. 

The electrical conductance at the Fermi level is shown in Fig.~\ref{fig:PD}(b). The system is an insulator except along the line separating phases with different Chern numbers. There, the system is conducting with conductance equal to half of conductance quantum $G_0/2$. 

The existence of the non-zero conductance and its connection to the phase diagram can be explained by inspecting the localization properties through the energy level statistics. According to the random matrix theory~\cite{Mehta04}, energy levels of localized states are distributed randomly inside an energy window following the Poisson distribution $P_\mathrm{P}(s)=e^{-s}$. The level-spacing variance $\langle s^2\rangle-\langle s\rangle^2$ is of the order of 1. Level statistics of extended states, on the other hand, follows the Wigner-Dyson distribution $P_\mathrm{WD}(s)=\frac{32s^2}{\pi^2}e^{-4s^2/\pi^2}$, with level-spacing variance equal to $0.178$. In Fig.~\ref{fig:PD}(c), we show level-spacing variance for a system at $u=-1$ and varying $W$ and energy $E$. 
It reveals the feature that distinguishes the strongly disordered regime from the clean and weakly disordered ones: the closing of the spectral gap and the topological transition occur at different parameter values. At $u=-1$, the spectral gap closes at $W_g\approx2.5$, when localized states (dark green) first appear at the Fermi energy, whereas the Chern number changes only when the mobility gap closes at $W_c\approx6$.  For $W>W_c$, there are no more extended states (light green) in the system, which  marks the topological Anderson phase transition to a trivial Anderson insulator. For $W_g<W<W_c$ the system is a topological insulator without a spectral gap. The in-gap states are localized and do not carry current, which is why they affect neither the Chern number nor the conductance. As shown in Sec.~\ref{sec:quenches}, they nevertheless govern the generation of excitations during a quench.

\section{Critical scaling}
\label{sec:crtitical}
To characterize the topological Anderson phase transition, we investigate the universal behaviour of conducting and topological properties and from them extract the correlation length $\xi$ and the relaxation time $\tau_r$. Close to the critical point, these are expected to be power-law functions of disorder strength
\begin{equation}
\xi(W)\propto|W-W_c|^{-\nu}, \qquad \tau_r(W)\propto \xi(W)^z\propto|W-W_c|^{-\nu z}.
\end{equation}
The power exponents $z$ and $\nu$ are termed the dynamical and the correlation-length critical exponent, respectively. Critical behaviour is important to understand also for systems out of equilibrium. For clean topological systems that are driven with a finite rate across the critical point, it was shown that the critical behaviour is imprinted in system's non-equilibrium properties through the KZ mechanism~~\cite{Damski05,Dutta10,Ulcakar2018,Ulcakar19}. We set $u=-1$ throughout the critical scaling analysis and quench dynamics. In this case, the system becomes a trivial insulator at $W_c=6.01\pm0.02$ as estimated from the maximum of electrical conductance.

\subsection{Correlation length critical exponent}
\label{sec:xi}
\begin{figure}
\includegraphics[width=0.33\textwidth]{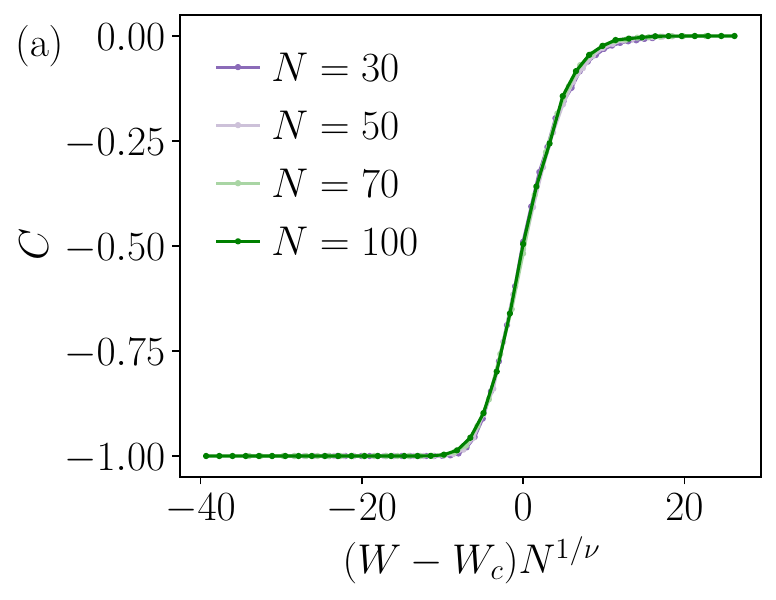}\includegraphics[width=0.31\textwidth]{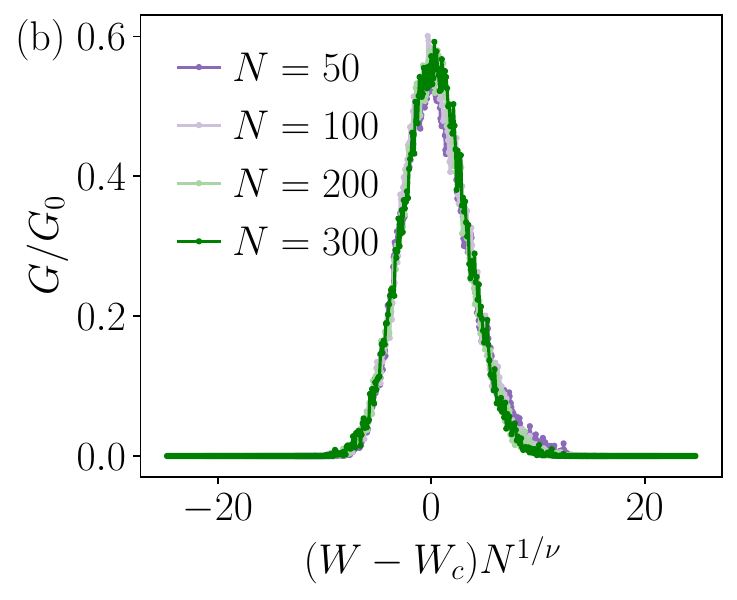}\includegraphics[width=0.31\textwidth]{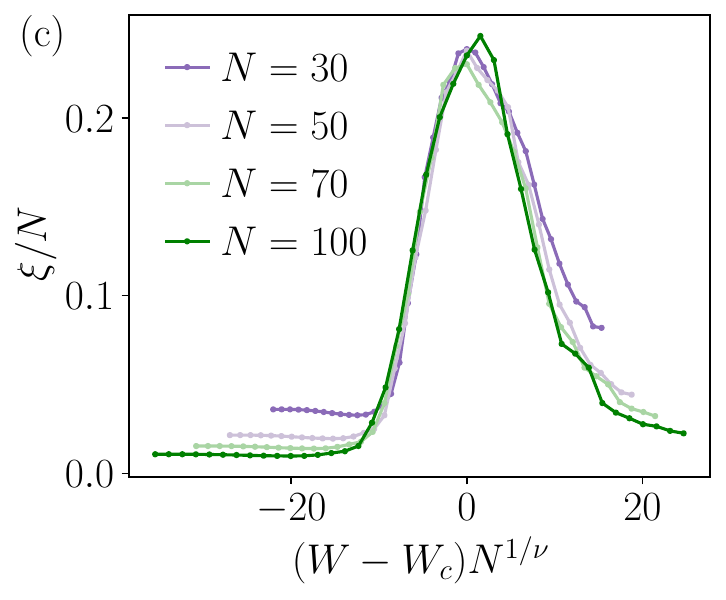}
	\caption{Finite-size scaling plots across the phase transition of (a) the Chern number, (b) the electrical conductance and (c) the correlation length $\xi/N$ estimated from the LCM profile. Quantities are presented as functions of the rescaled control parameter $(W-W_c)N^{1/\nu}$ for different system sizes.}
	\label{fig:xi_FS}
\end{figure}
The correlation-length critical exponent $\nu$ is inferred from finite-size scaling of quantities, whose critical behaviour is governed by the divergence of the correlation length.  In a finite system of linear size $N$, this divergence is cut off by $N$, such that observables depend on the ratio $N/\xi$. For observables that exhibit single-parameter scaling, this leads to the scaling form
\begin{equation}
f(W,N) = F(|W - W_c| N^{1/\nu}),
\end{equation}
where $F$ is a universal scaling function and $f$ is the considered quantity.
We identify this behaviour in both the Chern number and the linear electrical conductance, as shown in Fig.~\ref{fig:xi_FS}(a) and Fig.~\ref{fig:xi_FS}(b). By optimizing the collapse of the rescaled data with respect to the exponent $\nu$, we obtain $\nu=2.7\pm0.1$ from the conductance and $\nu=2.5\pm0.1$ from the Chern number. The optimal value is determined by minimizing the $\chi^2$ cost function across all combinations of system sizes, with the reported estimate given by the mean and the uncertainty by the corresponding standard deviation of the optimal values obtained from each size combination. 

\begin{figure}
\includegraphics[width=1\textwidth]{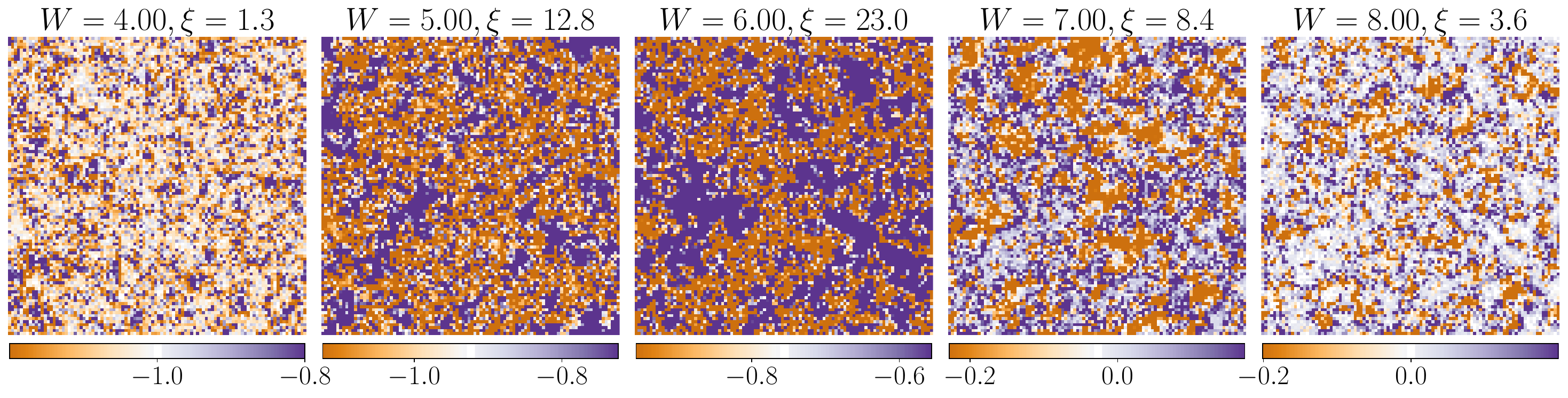}
	\caption{Real-space profile of the local Chern marker of the ground state at different disorder strengths that range across the phase transition.}
	\label{fig:lcm}
\end{figure}
In the following, we show that the critical length scale appears also in the real-space profile of the LCM. Its profile becomes highly inhomogeneous  as disorder strength is increased, see Fig.~\ref{fig:lcm}. System consists of regions, where the LCM deviates above and below the average value. The typical size of these regions is increased as the disorder strength approaches the critical point. We estimate the typical length scale of these regions from the autocovariance function of the LCM \cite{Ulcakar20,Favata25}, 
\begin{equation}
R(r)=\frac{\sum_{|\mathbf{r}|=r}\sum_{\mathbf{r'}}c(\mathbf{r'})c(\mathbf{r'}+\mathbf{r})}{\sum_{|\mathbf{r}|=r}\sum_{\mathbf{r'}}c(\mathbf{r'})^2}.
\end{equation}
We identify the typical length scale $\xi$ in the LCM as the distance at which the autocovariance function crosses zero, $R(\xi)=0$. The resulting disorder-averaged length scale as a function of rescaled disorder strength is shown in Fig.~\ref{fig:xi_FS}(c). By imposing the finite-size scaling ansatz and finding the best fit between curves for different system sizes, we find that $\nu=2.5\pm0.3$. This matches the exponent obtained from the conductance and the Chern number and confirms that the LCM exhibits the critical length scale at the topological Anderson phase transition.  The advantage of such formulation is that the length scale remains valid also out of equilibrium and allows for dynamical tracking of the system, as explored in Sec.~\ref{sec:quenches}.

\subsection{Dynamical critical exponent}
\begin{figure}
\centering
\includegraphics[width=0.32\textwidth]{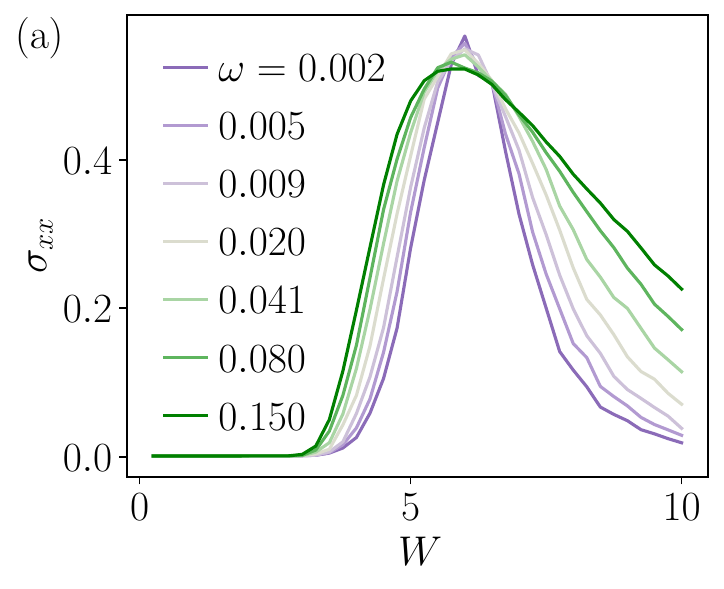}\includegraphics[width=0.32\textwidth]{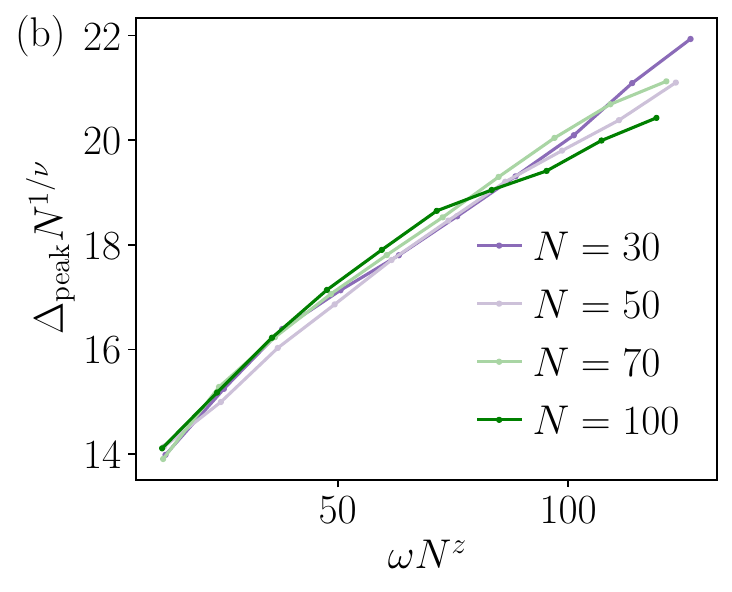}
	\caption{(a) Conductivity $\sigma_{xx}$ as a function of disorder strength at various $\omega$. (b) Full width at half maximum of $\sigma_{xx}$ as function of the rescaled frequency $\omega N^{z}$ for different system sizes.}
	\label{fig:sxx_FS}
\end{figure}
Dynamical scaling of the IQHE has been  measured from the critical behaviour of the conductance at a finite temperature or frequency of the driving electrical field~\cite{Wei88,Engel93,Huckelstein95}. A key feature, shared by both the IQHE and the disordered system considered here, is that the direct-current (DC) longitudinal conductivity vanishes away from the critical point, resulting in a singular peak at $W=W_c$ in the thermodynamic limit. At finite frequency, this singularity is broadened, giving rise to a peak in $\sigma_{xx}$ with a finite width. Fig.~\ref{fig:sxx_FS}(a) shows the conductivity as a function of disorder strength for different frequencies. The data clearly show the broadening of the conductivity peak as the frequency is increased. The finite width for frequencies tending to 0 is a consequence of the finite system size.

To characterize dynamical scaling, we analyze the scaling of the frequency-induced peak width. For a finite system, the conductivity depends on the ratio of the correlation length to the system size, $\xi/N$, and on the frequency through the dimensionless combination $\omega\tau_r$. Using the dynamical scaling relation $\tau_r\propto \xi^z$, the conductivity can be expressed with the scaling~\cite{Avishai96,Schweitzer03}
\begin{equation}
\sigma_{xx}(\omega,W)=G_0F(|W-W_c| N^{1/\nu}, \omega\,N^z).
\label{eq:sxxscaling}
\end{equation}
Here, $\omega\,N^z$ defines the dimensionless time scale associated with the critical spreading over the system of size $N$. From Eq.~\eqref{eq:sxxscaling}, the scaling of the peak width follows directly as
\begin{equation}
\Delta_\mathrm{peak}(\omega,N)=N^{-1/\nu}F(\omega N^z).
\end{equation}
The finite-size scaling of the peak width is shown in Fig.~\ref{fig:sxx_FS}(b). By optimizing the collapse of the rescaled curves as a function of $z$ for a fixed $\nu=2.5$ on the interval $\omega\in[\eta,9\eta)$, we obtain $z=1.96\pm0.07$. This is close to $z=2$, which is the value found for other Anderson-type phase transitions for non-interacting fermions in two dimensions, including the IQHE~\cite{Huckelstein95}. This places the disordered Chern insulators in the same universality class as the IQHE for non-interacting fermions.~\footnote{Experiments on IQHE found $z\approx1$ \cite{Evers08}, which is a consequence of electron interactions.}

\section{Quenches}
\label{sec:quenches}

\begin{figure}
\includegraphics[width=0.32\textwidth]{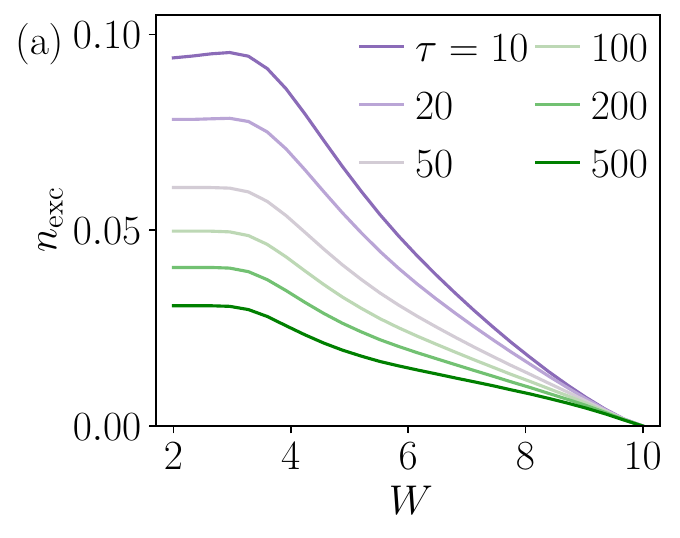}\includegraphics[width=0.32\textwidth]{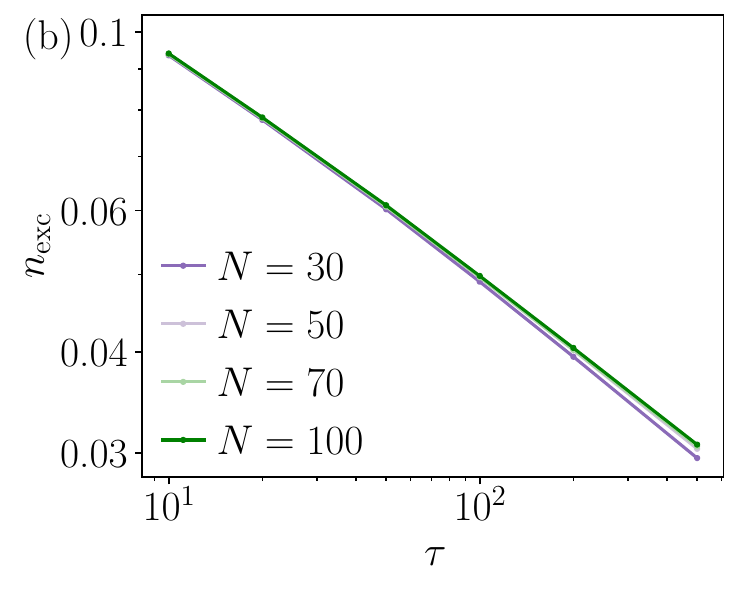}
\includegraphics[width=0.32\textwidth]{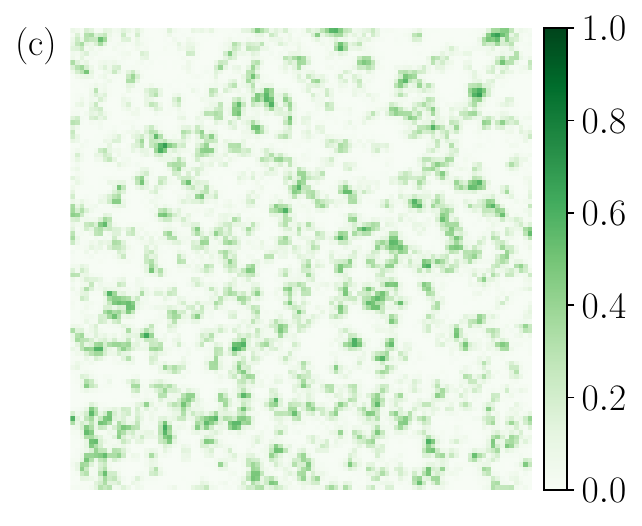} \\
\includegraphics[width=0.32\textwidth]{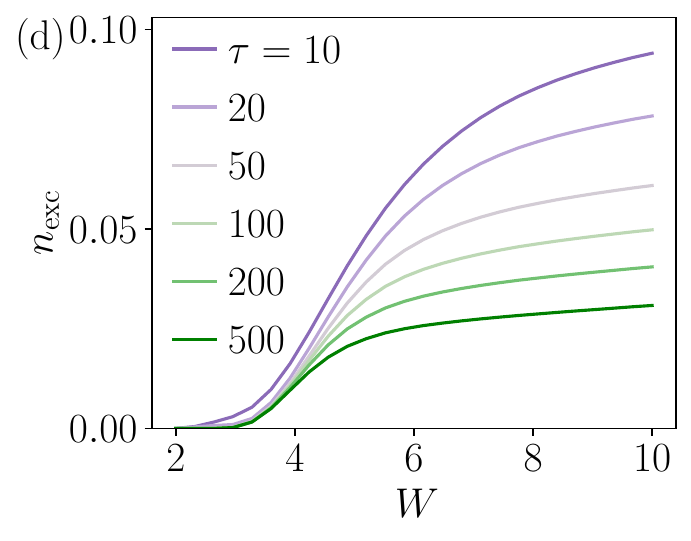}\includegraphics[width=0.32\textwidth]{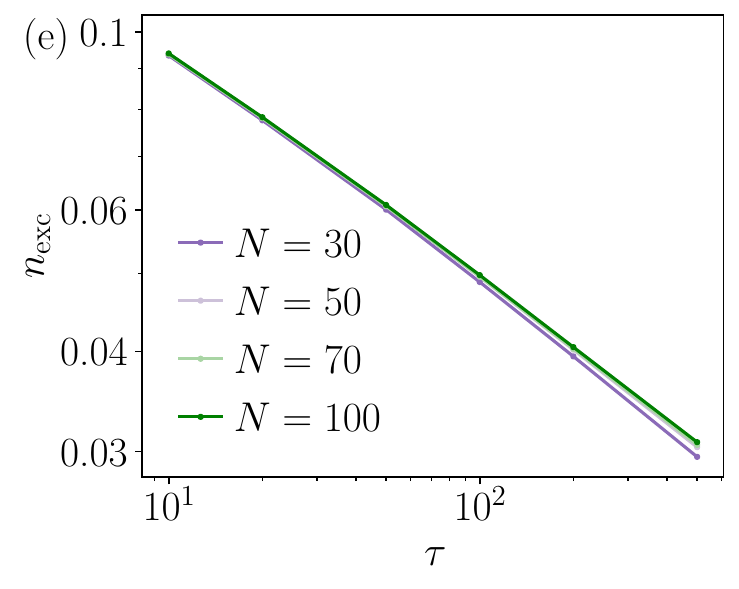}\,\,\includegraphics[width=0.32\textwidth]{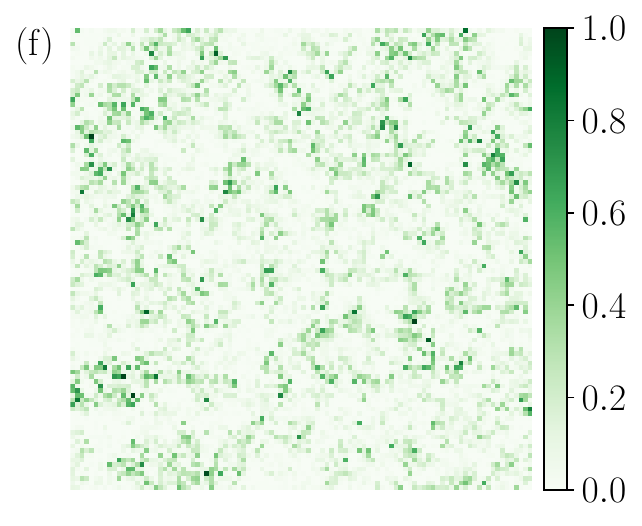}
\caption{Excitation density (a), (d) during quenches with different $\tau$ and (b), (e) at the end of the quench as a function of $\tau$. (c), (f) show spatial distribution of excitations after a quench with $\tau=10$ and $N=100$. (a), (b) and (c) correspond to the trivial-to-topological protocol and (d), (e) and (f) to the topological-to-trivial protocol.}
	\label{fig:Quench_exc}
\end{figure}
In the following, we study systems driven linearly in time across the phase transition and analyze the resulting non-equilibrium dynamics in relation to ground-state critical properties. We consider quenches starting from the ground state at $u=-1$, where the disorder strength is varied linearly over a time $\tau$ from an initial value $W_0$ to a final value $W_1$, $W(t)=W_{0}+(W_{1}-W_{0})t/\tau$. Disorder landscape remains unchanged. We consider quenches in both directions: from the topological phase ($W_0=2$) to the trivial phase ($W_1=10$), and vice versa. During the quench, the spectral gap is closed due to localization, which drives the system out of equilibrium.

To describe the resulting non-equilibrium state, we calculate the density of excitations, defined as
\begin{equation}
n_\mathrm{exc}(\mathbf{r},t)=\sum_\alpha\langle\mathbf{r}\alpha|\hat{P}(t)\hat{P}_c(t)|\mathbf{r}\alpha\rangle.
\end{equation}
Here, $\hat{P}(t)=\sum_n |\Psi_n(t)\rangle\langle \Psi_n(t)|$ is the projector onto the non-equilibrium occupied states and $\hat{P}_c(t)$ is the projector onto the unoccupied eigenstates of $H(t)$. Fig.~\ref{fig:Quench_exc}(a) and Fig.~\ref{fig:Quench_exc}(d) show the generation of excitations during quenches with different durations for both quench protocols. Due to the absence of the spectral gap for $W\gtrsim 2.5$, the time evolution is non-adiabatic, and excitations are generated continuously during the  quench. The excitation generation rate is approximately constant and depends on the quench duration $\tau$, leading to a $\tau$-dependent excitation density at the end of the quench. This is shown in Fig.~\ref{fig:Quench_exc}(b) and Fig.~\ref{fig:Quench_exc}(e) for different system sizes. The final excitation density follows a power-law scaling with $\tau$, with an exponent $0.281\pm0.004$ at $N=100$ for both directions of the quench. The real space distribution of excitations is shown in Fig.~\ref{fig:Quench_exc}(c) and Fig.~\ref{fig:Quench_exc}(f) for both considered protocols. The excitations are point-like due to the localization of eigenstates. In clean topological systems, it was shown that the excitation density follows the KZ scaling for defects with the power exponent $d\nu/(1+z\nu)=0.85\pm0.03$ ~\cite{Damski05,Dutta10,Ulcakar2018,Ulcakar19}, where the dimension $d=2$. Our results clearly show that the excitation generation is not described by the KZ mechanism.

\begin{figure}
\includegraphics[width=1\textwidth]{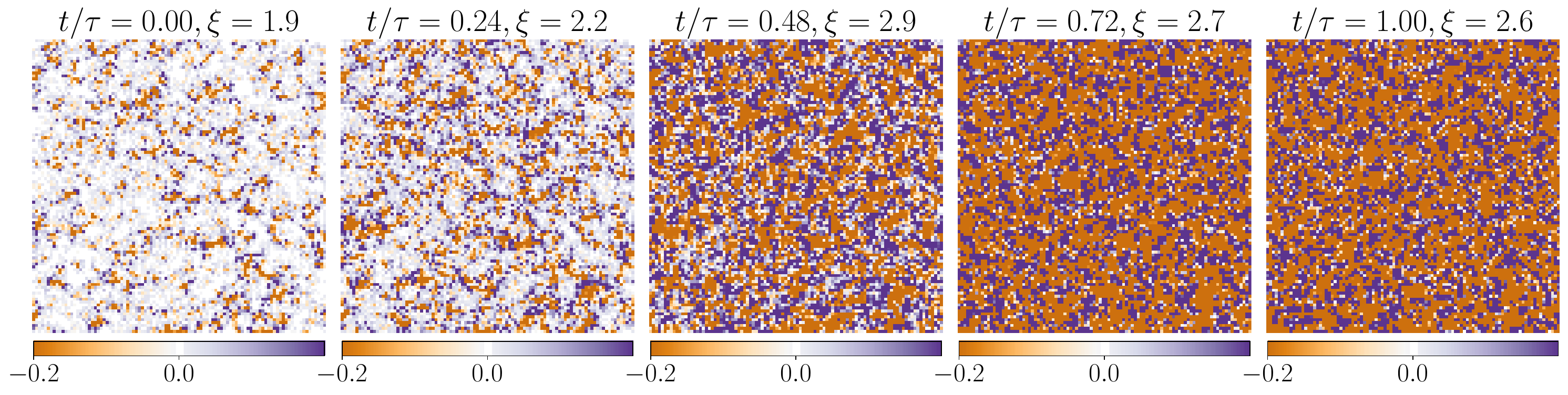}
	\caption{Time evolution of the local Chern marker shown at different time slices during the trivial-to-topological quench with $\tau=100$.}
	\label{fig:lcm_quench}
\end{figure}
The LCM can be calculated for the non-equilibrium state by inserting the projector $\hat{P}(t)$ in Eq.~\eqref{LCM}. Fig.~\ref{fig:lcm_quench} shows the real-space profile of the LCM at different times during a quench with $\tau=100$ from the trivial to the topological phase. As expected from Refs.~\cite{Caio2015,Ulcakar2018}, the spatial average of the LCM, the Chern number, remains invariant throughout the evolution~\cite{Rigol2015}~\footnote{For long $\tau$, the Chern number may change due to finite size effects.}. 
\begin{figure}[h]
\centering
\includegraphics[width=0.32\textwidth]{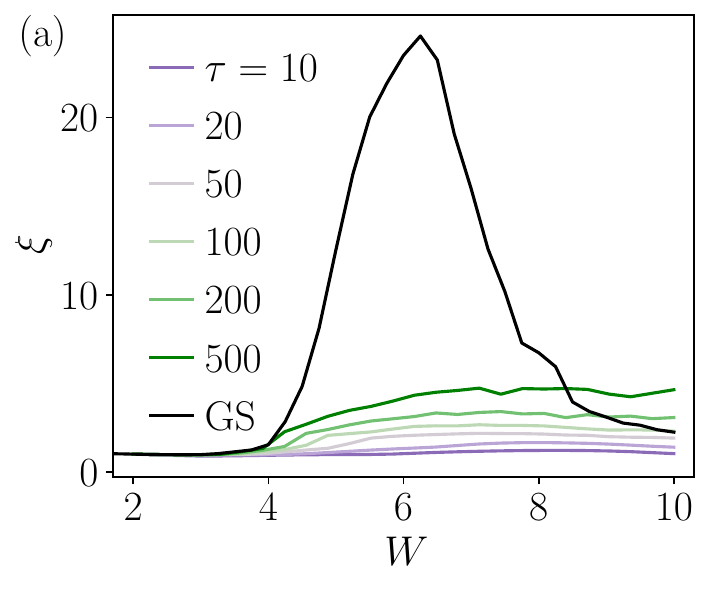}\includegraphics[width=0.32\textwidth]{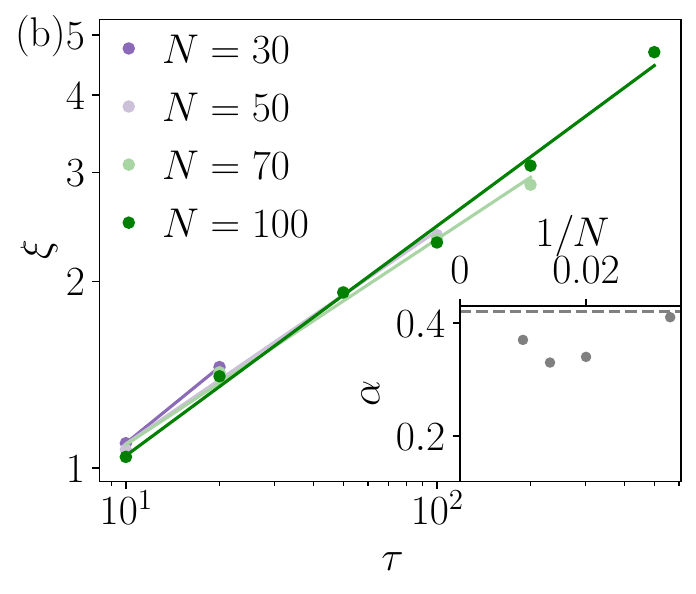}\\
\includegraphics[width=0.325\textwidth]{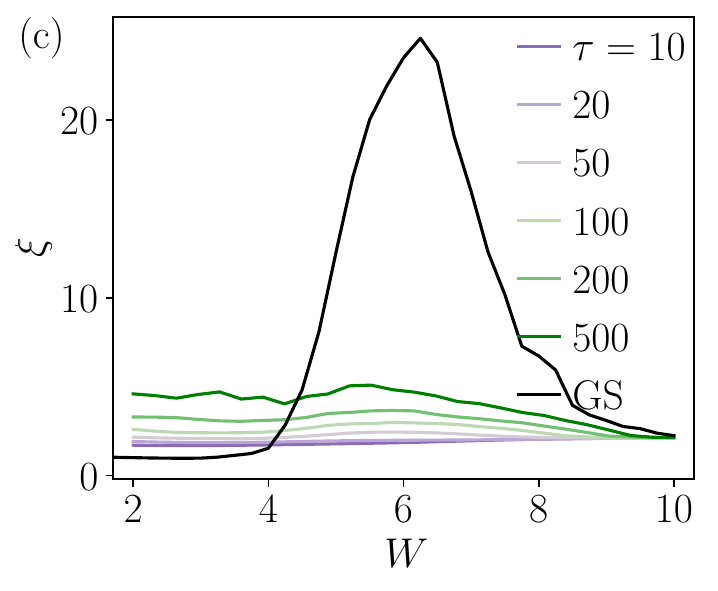}\includegraphics[width=0.32\textwidth]{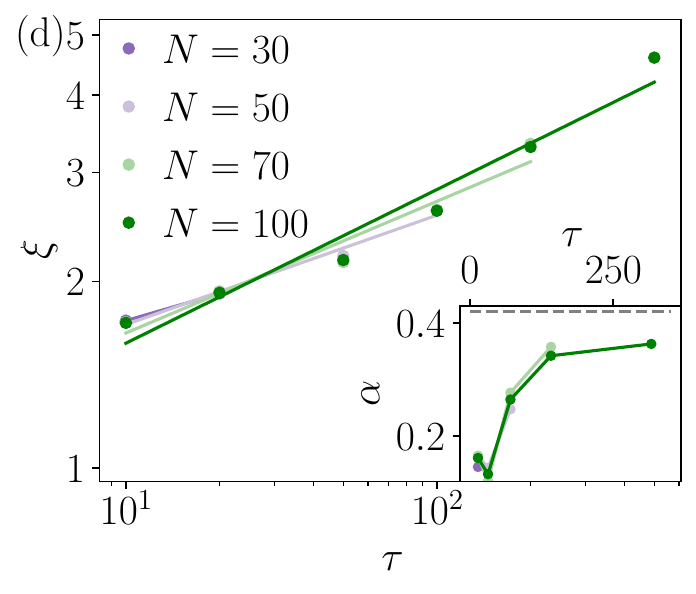}
\caption{Non-equilibrium  length scale extracted from the profile of the LCM  (a) during the topological-to-trivial quench and (c) during the trivial-to-topological quench for different quench durations. The black curve denotes the length scale in the ground state. The scaling of the final length scale with $\tau$ (b) for the topological-to-trivial quench and (d) for the trivial-to-topological quench.  The inset in (d) shows the effective power-law exponent as a function of $\tau$ for different system sizes. The inset in (b) shows the finite-size scaling of the power-law exponent evaluated over the whole range of $\tau$.}
	\label{fig:Quench_xi}
\end{figure}

We first analyse the non-equilibrium length scale during the topological-to-trivial quench. As shown in Fig.~\ref{fig:Quench_xi}(a), it follows the adiabatic curve (black) until $W\approx4$,  beyond which it starts to deviate, exhibiting the characteristic freeze-out behaviour. The dependence of this length scale, calculated at the end of the quench, on the quench duration is shown in Fig.~\ref{fig:Quench_xi}(b)~\footnote{We omit the data where due to finite size effects the Chern number changes for more than $0.1$.}.  It exhibits a power-law scaling $\tau^\alpha$ with $\alpha=0.37\pm0.03$ for $N=100$, which is close to the KZ prediction $\alpha=\nu/(1+z\nu)=0.42\pm0.01$. The values for smaller systems fluctuate (see inset to Fig.~\ref{fig:Quench_xi}(b)), however they are all quite close to the KZ prediction. 

For the trivial-to-topological quench, the applicability of the KZ mechanism is less evident. As shown in Fig.~\ref{fig:Quench_xi}(c), the non-equilibrium length scale deviates from its adiabatic behaviour early in the quench, without exhibiting a clear freeze-out regime. Consistently, its final value does not follow the power-law ansatz over the full range of quench durations, as illustrated by the linear fits in Fig.~\ref{fig:Quench_xi}(d). To investigate this deviation, we estimate an effective exponent $\alpha$ as a function of $\tau$ from pairs of consecutive quench times. As shown in the inset of Fig.~\ref{fig:Quench_xi}(d), the effective exponent moves toward the KZ prediction as $\tau$ increases. This trend suggests that the deviations from the KZ scaling at shorter quench times may arise because the freeze-out occurs outside the critical regime in which universal scaling is expected. Reaching the asymptotic KZ regime may therefore require substantially longer quench times. However, reliably accessing this regime would also require larger system sizes to control finite-size effects, which limit the reliability of the extracted scaling at long quench times.
We nevertheless present additional results for longer quenches in Appendix~\ref{app:longtau}, focusing on time scales before the finite-size effects. These show that the non-equilibrium length scale exhibits clearer freeze-out behaviour before finite-size effects become significant, providing further evidence for the onset of KZ scaling at longer durations.

We attribute the asymmetry between the two quench directions  to the asymmetry of the ground-state properties around the critical point. Since the disorder strength is varied at fixed $u$, the quench path does not cross the phase boundary perpendicularly  (see Fig.~\ref{fig:PD}(a)). Consequently, the ground-state length scales differ. Specifically the length scale  at $W=10$ is approximately twice that at $W=2$, making the power-law growth of the length scale more difficult to resolve for the trivial-to-topological quench at the accessible quench durations. 

While we observe signatures of the characteristic KZ freeze-out behavior, it is important to note that our results also show deviations from the standard KZ picture. When the quench is from the topological to the trivial side, we do see signatures of the freeze-out behaviour with characteristic growth of the LCM length scale with expected exponent. However, the freeze-out is not connected to the adiabaticity. For the  quench with $\tau=500$, the excitations in Fig.~\ref{fig:Quench_exc}(d) appear  already at $W\approx2.5$ where the spectral gap closes, whereas the LCM length scale in Fig.~\ref{fig:Quench_xi}(c) follows the ground state one until a much larger $W\approx4.2$. This suggests that the LCM length scale is related to the topological phase transition while the excitation generation is related to the closing of the spectral gap (see Fig.~\ref{fig:PD}(c)).

\section{Conclusion}
\label{sec:conclusion}
We investigated  the ground-state critical properties and the quench dynamics of strongly disordered Chern insulators. The disorder-driven phase transition occurs when the mobility gap closes, driving the system into a trivial Anderson insulating phase. This transition is signaled by a non-zero electrical conductivity and a change in the Chern number. The associated critical length scale manifests directly as the characteristic size of inhomogeneities in the real-space profile of the LCM. We extracted the critical exponents of the topological Anderson transition, $z=2$ and $\nu\approx2.5$,  which fits the same universality class as the non-interacting IQHE. In contrast, experiments on the IQHE typically report a dynamical exponent $z=1$, which is a consequence of electron–electron interactions. It would therefore be interesting to investigate whether a similar renormalization of $z$ arises in disordered Chern insulators in the presence of interactions.

We further studied the non-equilibrium dynamics induced by quenches of the disorder strength across the phase transition. Unlike in the clean and weakly disordered Chern insulators, strong disorder separates the closing of the spectral gap and the mobility gap, where only the latter leads to the topological transition. We found that the time evolution remains strongly non-adiabatic throughout the entire quench due to the absence of a spectral gap. Consequently, unlike in clean topological systems, the dynamics  does not follow the conventional KZ scaling.  The excitation density does not exhibit the freeze-out behaviour, and it depends only weakly on the quench duration, indicating that the excitation dynamics is not governed by the equilibrium critical scaling. In contrast, the quench-induced LCM length scale exhibits the freeze-out behaviour. For the topological-to-trivial quench, its scaling is close to the KZ predictions. For the reverse direction, the length scale  approaches the KZ prediction in the limit of long quenches. 
While the KZ behaviour cannot be explained by the excitation dynamics alone, the freeze-out of the LCM has to take the excitations into account. This suggest that are two types of excitations, where one directly influences the fluctuations of the LCM while the other one does not. The distinction between these two types of states and their relation to the LCM provides an interesting question for further studies.

Another interesting direction for future work would be to extend this analysis to disordered topological systems in other symmetry classes, where different critical behaviour may emerge. In particular, disordered topological superconductors provide a promising platform, as they can host an  metallic phase (the so-called Majorana metal) \cite{Chalker2001,Medvedyeva10,Zakharov25}.

\section*{Acknowledgments}
We thank L. Vidmar and J. C. Budich for interesting discussions and H. Torbatiyan for helpful comments. 

\paragraph{Funding information}
This work was supported by the Slovenian Research and Innovation Agency (ARIS) under contracts no. P1-0044 and J1-50005. G. Lemut acknowledges support by the
Deutsche Forschungsgemeinschaft (DFG, German Research Foundation) as part of the German Excellence
Strategy - EXC3112/1 - 533767171 (Center for Chiral Electronics).

\appendix

\section{Numerical calculation of the longitudinal conductivity}

\subsection{Velocity operator}
The longitudinal conductivity $\sigma_{xx}(\omega)$ is calculated using the Kubo--Greenwood formula. For each disorder realization and disorder strength $W$, we first construct the single-particle Hamiltonian $\hat H$ and obtain its eigenvalues and eigenvectors using the Hermitian eigensolver \texttt{scipy.linalg.eigh}.

The velocity operator is obtained from the position operator according to $\hat v_x = i[\hat X,\hat H]/\hbar$, which in the site-orbital basis gives 
\begin{equation}
(\hat v_x)_{i\alpha,j\beta}
=
i\Delta x_{i\alpha,j\beta}(\hat H)_{i\alpha,j\beta}/\hbar.
\label{eq:vx_commutator}
\end{equation}
Since periodic boundary conditions are used in the $x$ direction, the displacement $x_{i\alpha}-x_{j\beta}$ is equal to 
\begin{equation}
\Delta x_{i\alpha,j\beta}
=
\left[
(x_{i\alpha}-x_{j\beta})+\frac{N}{2}
\right]\mathrm{mod}N
-
\frac{N}{2},
\label{eq:minimum_image}
\end{equation}
where $N$ is the linear system size. After diagonalisation of the Hamiltonian, the velocity operator is transformed to the eigenbasis,
\begin{equation}
(\hat v_x)_{nm}=\langle \Psi_m|\hat{v}_x|\Psi_n\rangle.
\end{equation}

\subsection{Kubo--Greenwood conductivity}

\begin{figure}[h]
\centering
\includegraphics[width=0.34\textwidth]{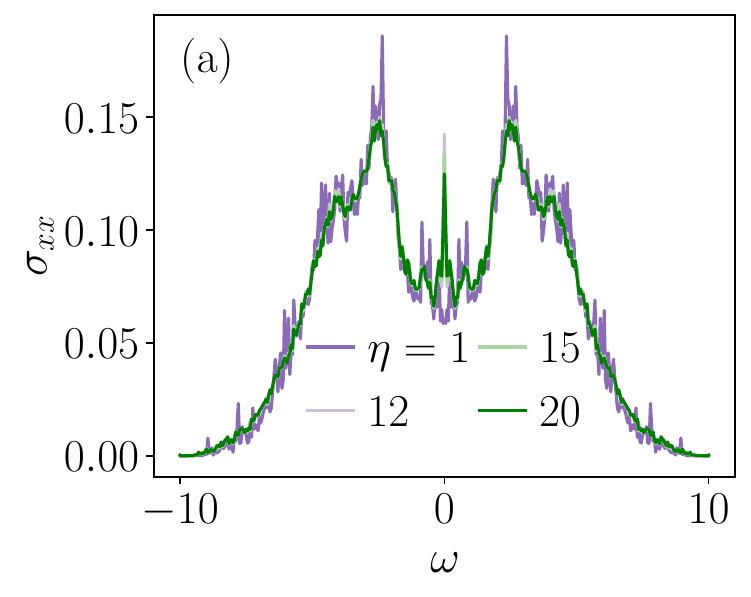}
\includegraphics[width=0.32\textwidth]{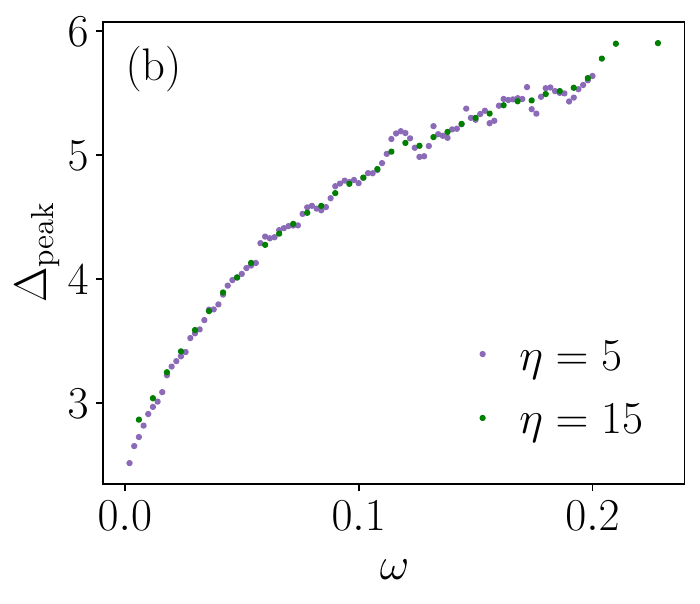}
	\caption{Analysis of the broadening parameter $\eta$. (a) AC conductivity $\sigma_{xx}$ and (b)  full width at half maximum of $\sigma_{xx}$ in disorder strength $W$ as a function of the frequency $\omega$, for different values of $\eta$. System size was set to $N=50$.}
	\label{fig:sxx_eta}
\end{figure}

The frequency-dependent longitudinal conductivity is calculated from
\begin{equation}
\sigma_{xx}(\omega)=-\frac{e^2\hbar}{N^2}\sum_{m\neq n}\frac{f(E_m)-f(E_n)}{E_m-E_n}|\langle \Psi_m|\hat{v}_x|\Psi_n\rangle|^2\delta(E_n-E_m-\hbar\omega).
\end{equation}
where $f(E_n)$ is the Fermi--Dirac occupation of state at energy $E_n$. Throughout this work we consider zero temperature and half filling. 
The Dirac delta function is replaced by a Lorentzian broadening,
\begin{equation}
\delta(x)=\frac{1}{\pi}\frac{\eta}{x^2+\eta^2},
\end{equation}
where $\eta$ is the broadening parameter. In the main text, we present results for $\eta=15/N^2$. The prefactor $15$ corresponds to a representative bandwidth for the range of disorder strengths $W$ considered, while the factor $1/N^2$ follows the scaling of the mean single-particle energy-level spacing with system size. The broadening therefore sets the effective energy resolution of the conductivity calculation. We have verified that the calculated $\sigma_{xx}$ is converged for this choice of $\eta$, as demonstrated in Fig.~\ref{fig:sxx_eta}.

\section{Method for time evolution}

We study the non-equilibrium dynamics of the disordered QWZ model by directly integrating the time-dependent Schr\"{o}dinger equation, 
where the time dependence of the Hamiltonian originates from the time-dependent disorder strength $W(t)=W_1+(W_2-W_1)t/\tau$, where $\tau$ is the quench time. The disorder realization is generated before the time evolution and remains fixed throughout each simulation.

The time evolution is performed using the Crank--Nicolson (CN) scheme. It discretizes the time-dependent Schr\"{o}dinger equation as into time-steps $\Delta t$ as:
\begin{equation}
\left[
I+\frac{i\Delta t}{2}\hat H(t)
\right]\ket{\psi(t+\Delta t)}
=
\left[
I-\frac{i\Delta t}{2}\hat H(t)
\right]\ket{\psi(t)}.
\label{eq:CN}
\end{equation}
Here, we set $\hbar=1$. Equivalently, the propagation can be written as
\begin{equation}
\ket{\psi(t+\Delta t)}
=
\hat U_{\rm CN}(t,\Delta t)\ket{\psi(t)},
\end{equation}
where
\begin{equation}
\hat U_{\rm CN}(t,\Delta t)
=
\left[
I+\frac{i\Delta t}{2}\hat H(t)
\right]^{-1}
\left[
I-\frac{i\Delta t}{2}\hat H(t)
\right].
\label{eq:cayley}
\end{equation}
For a Hermitian Hamiltonian, the propagator $\hat U_\mathrm{CN}$ is unitary, such that the norm of the wave function is preserved up to numerical round-off errors. 

In our implementation, the Hamiltonian is represented as a sparse matrix of dimension $2N^2$. At every time step, we update $\hat H(t)$ and calculate $ \ket{\psi(t+\Delta t)}$ by solving the linear system from Eq.~\ref{eq:CN} using the sparse direct solver \texttt{scipy.sparse.linalg.spsolve}. The time step used in the calculations is $\Delta t=0.2$, which fulfills the convergence requirements. The initial many-body state is represented by the occupied single-particle wave functions of the initial Hamiltonian. The time-evolution Eq.~\eqref{eq:CN} is applied to the occupied single-particle states independently. 
As the Hilbert space dimension scales as 2$N^2$, increasing the linear system size rapidly increases both the memory requirements and the cost of the sparse linear solves. The largest systems considered in this work contain $N=100$ lattice sites.

\section{Evaluation of critical exponents and the estimates of errors}

To quantify the sensitivity of the extracted exponents to finite-size
effects, we performed the collapse separately for every pair of system sizes.
For a trial exponent, the data for each pair were expressed in terms of the
corresponding rescaled variable $(W-W_c)N^{1/\nu}$, linearly interpolated, and compared within
the common interval of the two curves.  We then minimized a mean-square
mismatch between the two rescaled curves.  A quadratic fit to the mismatch as
a function of the trial exponent was used to locate its minimum for each size
pair.

The value quoted in the main text is the mean of the six pairwise estimates from different pairs of system sizes,
while its uncertainty is their standard deviation.  We interpret the largest contribution to this spread
as a finite-size systematic uncertainty, rather than as the statistical error
with which the minimum for a single pair can be located.

For the correlation-length exponent, the Chern-number collapse gives
\begin{equation}
  \nu_C=2.451\pm0.107,
  \qquad 2.272\leq\nu_{C,ij}\leq2.614,
\end{equation}
whereas the conductance collapse gives
\begin{equation}
  \nu_G=2.688\pm0.109,
  \qquad 2.514\leq\nu_{G,ij}\leq2.839.
\end{equation}
The two estimates are obtained from different observables and different sets
of system sizes.  Their finite-size drifts need not be identical, and their
full pairwise ranges overlap.  We therefore retain the rounded values
$\nu_C=2.5\pm0.1$ and $\nu_G=2.7\pm0.1$ and interpret them as consistent
within the resolution of the available sizes.

For the dynamical exponent, the six size-pair estimates are
\begin{equation}
  1.874,\quad 1.957,\quad 2.075,\quad
  1.938,\quad 1.995,\quad 1.904,
\end{equation}
yielding $z=1.96\pm0.07$ and the full range
$1.87\leq z\leq2.08$.  The collapse also depends on the
correlation-length exponent used to rescale the peak width.  Repeating the
analysis for $\nu=2.4$, $2.5$, and $2.6$ gives
\begin{equation}
  z=2.049,\quad 1.957,\quad 1.881,
\end{equation}
respectively.  Combining the size-pair spread with the variation caused by
$\nu=2.5\pm0.1$ gives
\begin{equation}
  z=1.96^{+0.11}_{-0.10},
\end{equation}
which we quote as $z=2.0\pm0.1$.  This result expresses consistency with
$z=2$ within the finite-size resolution.

\section{Slow quenches from the trivial to the topological phase}
\label{app:longtau}
\begin{figure}[h]
\centering
\includegraphics[width=0.32\textwidth]{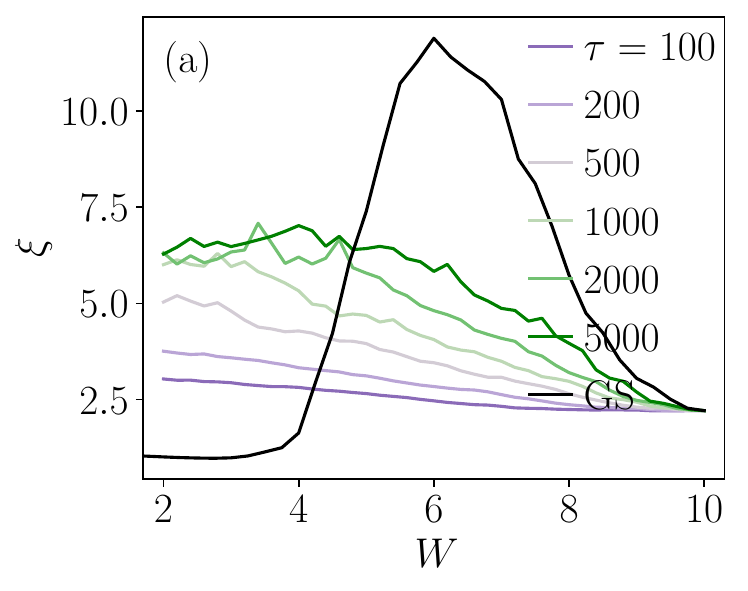}
\includegraphics[width=0.33\textwidth]{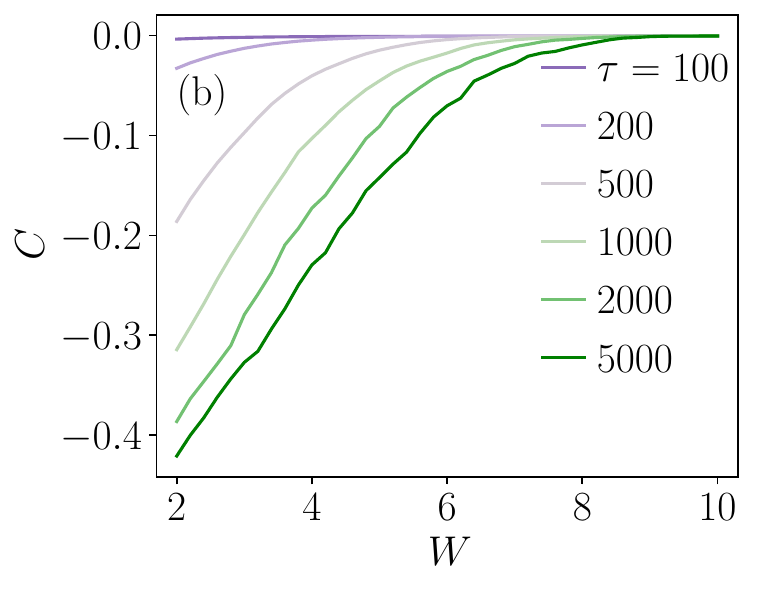}
\includegraphics[width=0.3\textwidth]{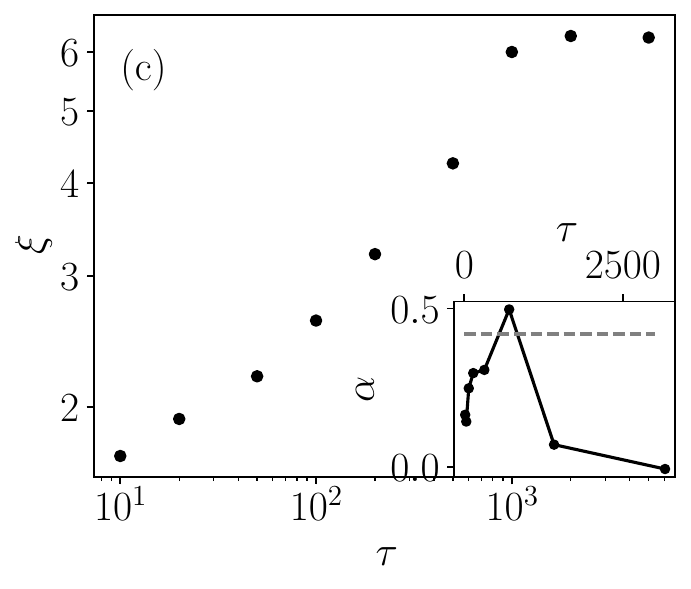}
\caption{(a) non-equilibrium length scale extracted from the LCM profile during trivial-to-topological quenches with different durations. The black curve shows the corresponding ground-state length scale. (b) Evolution of the Chern number during corresponding quenches. (c) Final non-equilibrium length scale as a function of the quench duration $\tau$. The inset shows the effective power-law exponent as a function of $\tau$. The system size is $N=50$.}
\label{fig:Quench_xi_longtau}
\end{figure}

In the main manuscript, we consider quench durations up to $\tau=500$, since longer evolution times lead to finite-size effects that compromise the reliability of the results. These effects can be monitored, for example, through changes in the Chern number, which is expected to remain invariant during unitary time evolution. However, the quenches from the trivial to the topological phase considered in the main manuscript are too short to reveal clear freeze-out behaviour characteristic of the Kibble--Zurek mechanism. Here, we  extend the analysis to longer quench durations, up to $\tau=5000$.

Fig.~\ref{fig:Quench_xi_longtau}(a) shows the non-equilibrium length scale extracted from the LCM profile during trivial-to-topological quenches for different durations. As $\tau$ increases, the non-equilibrium length scale follows the ground-state curve over a progressively larger range of disorder strengths, indicating the emergence of the freeze-out behaviour. The reliability of these results can be assessed by monitoring the Chern number, shown in Fig.~\ref{fig:Quench_xi_longtau}(b). While the Chern number remains conserved throughout the quench for $\tau=100$, deviations appear at longer durations. For $\tau=5000$, for example, finite-size effects become apparent around $W\approx8$. Nevertheless, the freeze-out behaviour emerges before these effects become significant.

Fig.~\ref{fig:Quench_xi_longtau}(c) shows the final non-equilibrium length scale as a function of the quench duration. Due to finite-size effects, the scaling can be reliably extracted only for $\tau<200$ at system size $N=50$.

\bibliographystyle{apsrev4-2}
\bibliography{bibliography_list}

\end{document}